# Design of a 60.8 K superconducting hydride LiMgZr$_2$H$_{12}$ at ambient pressure via Lithium doping

Qun Wei[a,*], Xinyu Wang[a], Jing Luo[a], Meiguang Zhang[b,*], Bing Wei[a,*]

[a]School of Physics, Xidian University, Xi'an 710071, China

[b]College of Physics and Optoelectronic Technology, Baoji University of Arts and Sciences, Baoji 721016, China

**ABSTRACT:** High-pressure hydrogen-rich compounds have long been regarded as promising room-temperature superconductor candidates; however, their practical applications are limited by their reliance on extreme compression. This study explores hydrogen-rich superconductors that may be stable at ambient pressures. Inspired by recent investigations of the MgZrH$_{2n}$ family, the LiMgZr$_2$H$_{12}$ structure with a *Pmmm* symmetry was constructed, and its thermodynamic, mechanical, and dynamical stability were evaluated using first-principles calculations. Electron–phonon coupling (EPC) analysis suggests that LiMgZr$_2$H$_{12}$ reaches a superconducting critical temperature ($T_c$) of 60.8 K at ambient pressure. Compared with MgZrH$_6$, Li doping significantly increases the contribution of hydrogen atoms to the electron density of states near the Fermi level ($E_F$) and enhances the EPC constant ($\lambda$) of the LiMgZr$_2$H$_{12}$ structure. LiMgZr$_2$H$_{12}$ exhibits a superconducting figure of merit of 1.56, which is significantly greater than that of MgZrH$_6$, demonstrating its outstanding potential for practical applications. This work guides ambient-pressure design of high-$T_c$ hydrides.

Superconductors have long attracted significant interest for their broad application potential, and the pursuit of high-temperature or even room-temperature superconductivity remains a key research focus in condensed matter physics and materials science. Over the past decade, breakthroughs have been achieved in studies on hydrogen-rich superconductors. Binary hydrides including H$_3$S [1], CaH$_6$ [2], LaH$_{10}$ [3], YH$_9$ [4], and YH$_6$ [5] exceed the conventional low-temperature superconductivity regime, achieving superconducting critical temperatures ($T_c$) above 200 K. The successful theoretical and experimental identification of these binary hydrides has further accelerated the development of high-pressure hydrides. Remarkably, several ternary hydrides predicted in recent years, such as $Li_2MgH_{16}$ [6], $Li_2NaH_{17}$ [7], $Li_2Na_3H_{23}$ [7], and $LaSc_2H_{24}$ [8], are theoretically expected to exhibit high $T_c$ values approaching or even exceeding room temperature under high pressures. However, most hydride superconductors are stable only above 150 GPa, severely limiting practical applications, making the realization of high-temperature superconductivity at low or ambient pressure a critical challenge.

Doping is one of the most effective strategies for tuning material properties and has been extensively demonstrated in multicomponent superconducting hydride systems [9,10]. Several ternary hydrides, including $(La,Ce)H_9$ [11], $LaBeH_8$ [12], $(La,Al)H_{10}$ [13], and $Li_2MgH_{16}$ [6], exhibit enhanced superconducting properties compared with their parent phases through elemental doping of the host framework. Both theoretical and experimental studies have demonstrated that, elemental doping can substantially reduce the pressure required for structural stabilization and increase the $T_c$.

Among the various candidates proposed for reducing the required pressure, the Mg–Zr–H system composed of the light element Mg and the transition metal Zr has attracted our attention. Within the $MgZrH_{2n}$ series, the $Pm\bar{3}$-$MgZrH_6$ phase is estimated, based on the Gor'kov–Kresin equation, to exhibit a $T_c$ value of 80.3 K at 36 GPa and a superconducting figure of merit ($S$) of 1.51, demonstrating the excellent superconducting potential [14]. Motivated by the excellent superconducting performance of $MgZrH_6$, we propose constructing a supercell and introducing the light element Li to reduce the pressure required for structural stabilization, thereby further enhance its superconducting properties. On the one hand, the atomic radius of Li is comparable to that of Mg, whereas its lower electronegativity allows for greater charge transfer from Li to H. On the other hand, the low atomic mass of Li can reduce the effective lattice mass and increase the logarithmic average phonon frequency $\omega_{\log}$. Accordingly, we construct a $1 \times 1 \times 2$ $MgZrH_6$ supercell along the lattice directions to obtain a $Mg_2Zr_2H_{12}$ supercell and then substitute one Mg atom with Li, ultimately yielding a $LiMgZr_2H_{12}$ quaternary hydride. In this study, we evaluate the thermodynamic, mechanical, and dynamical stability of the $LiMgZr_2H_{12}$ structure and systematically evaluate its superconducting properties via electron–phonon coupling (EPC) calculations. Our study provides useful guidance for future theoretical and experimental explorations of ambient-pressure high-temperature superconductors.

Geometry optimizations and related property calculations for LiMgZr$_2$H$_{12}$ were performed within the framework of density functional theory (DFT) using the Perdew–Burke–Ernzerhof (PBE) parametrization of the generalized gradient approximation (GGA) [15], as implemented in the Vienna *Ab initio* Simulation Package (VASP) [16]. The electron-ion interaction is described by projector augmented wave potential [17]. A plane-wave cutoff energy of 600 eV was employed, and the Brillouin zone was sampled using a Monkhorst–Pack *k*-point mesh [18] corresponding to a grid spacing of $2\pi \times 0.02$ Å$^{-1}$, ensuring that the total energy was converged to within $1 \times 10^{-5}$ eV/atom. The single-crystal elastic constants were obtained by fitting linear stress–strain relations [19]. Dynamic stability was evaluated using the finite displacement approach, and phonon spectra were calculated using the PHONOPY package [20]. The crystal orbital Hamiltonian population (COHP) and the corresponding integrated COHP (ICOHP) values were obtained using the LOBSTER code [21,22]. EPC was calculated using the Quantum ESPRESSO package [23] with a plane-wave kinetic energy cutoff of 60 Ry. For Brillouin-zone sampling, a $24 \times 24 \times 12$ *k*-point mesh together with a Gaussian smearing of 0.05 Ry was adopted to achieve convergence, while a $6 \times 6 \times 3$ *q*-point mesh was employed to compute the EPC constant. The superconducting critical temperature was estimated using the Allen–Dynes modified McMillan equation [24].

Based on the MgZrH$_6$ prototype, we constructed a LiMgZr$_2$H$_{12}$ structure with *Pmmm* symmetry and performed computational analyses of its thermodynamic, mechanical, and dynamical stability. The thermodynamic stability of the LiMgZr$_2$H$_{12}$ structure can be evaluated in terms of its formation energy, defined as follows [25,26]:

$$\Delta H = \frac{E(\text{LiMgZr}_2\text{H}_{12}) - E(\text{Li}) - E(\text{Mg}) - 2E(\text{Zr}) - 12E(\text{H})}{16} \quad (1)$$

where $E(\text{LiMgZr}_2\text{H}_{12})$ denotes the total energy of the compound, and $E(\text{Li})$, $E(\text{Mg})$, $E(\text{Zr})$, and $E(\text{H})$ denote the average energies of single Li, Mg, Zr, and H atoms in the crystals, respectively. The calculated formation energy of the LiMgZr$_2$H$_{12}$ structure is negative, and its elastic constants satisfy the Born stability criterion [27]. These results confirm that the structure is both thermodynamically and mechanically stable. Subsequently, we evaluated the dynamic stability of LiMgZr$_2$H$_{12}$ at ambient pressure by calculating its phonon dispersion. Fig. 2 shows that all phonon modes in the Brillouin zone exhibit positive frequencies, demonstrating that the LiMgZr$_2$H$_{12}$ structure is dynamically stable under ambient conditions. The LiMgZr$_2$H$_{12}$ structure is composed of H$_{12}$ cages centered by Zr, Li, and Mg atoms (Fig. 1). Each H$_{12}$ cage consists of 12 isosceles and 8 scalene triangles. LiMgZr$_2$H$_{12}$ exhibits a minimum H–H distance of 1.76 Å, which is substantially longer than the standard H–H covalent bond length in molecular H$_2$ (0.74 Å) at ambient pressure. Table 1 summarizes the detailed structural parameters of LiMgZr$_2$H$_{12}$ for further analysis.

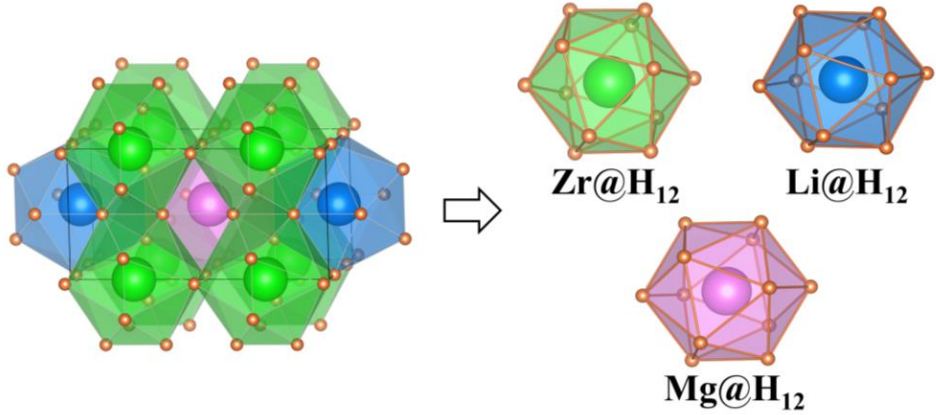

**Fig. 1.** Crystal structure of the quaternary hydride LiMgZr$_2$H$_{12}$ with *Pmmm* symmetry.

**Table 1**

Structural parameters of LiMgZr$_2$H$_{12}$ under ambient pressure.

| Phase | Lattice parameter (Å) | Wyckoff position | | | |
|---|---|---|---|---|---|
| | | Atoms | $x$ | $y$ | $z$ |
| *Pmmm* | $a = 3.7833$ $b = 3.7853$ $c = 7.5389$ $\alpha = \beta = \gamma = 90°$ | Li (1$f$) | 0.500 | 0.500 | 0 |
| | | Mg (1$h$) | 0.500 | 0.500 | 0.500 |
| | | Zr (2$q$) | 0 | 0 | 0.245 |
| | | H1 (2$r$) | 0 | 0.500 | 0.130 |
| | | H2 (2$r$) | 0 | 0.500 | 0.636 |
| | | H3 (4$v$) | 0.500 | 0.766 | 0.252 |
| | | H4 (2$i$) | 0.762 | 0 | 0 |
| | | H5 (2$j$) | 0.768 | 0 | 0.5 |

To investigate potential superconductivity, we calculated the phonon dispersion relations, projected phonon density of states (PHDOS), Eliashberg spectral function $\alpha^2F(\omega)$, and the EPC constant $\lambda$ for LiMgZr$_2$H$_{12}$ at ambient pressure. The corresponding results are summarized in Fig. 2. EPC analysis reveals that LiMgZr$_2$H$_{12}$ has an EPC constant $\lambda = 2.22$, which is significantly greater than that of MgB$_2$ at ambient pressure ($\lambda = 0.61$) [28] and that of MgZrH$_6$ at 36 GPa ($\lambda = 1.13$) [14], indicating the substantial EPC strength of LiMgZr$_2$H$_{12}$. Because of their relatively large atomic masses, Li, Zr, and Mg atoms are primarily associated with low-frequency phonon modes, whereas lighter H atoms dominate intermediate- and high-frequency phonon modes. In the low-frequency range (0–9 THz), the phonon modes primarily arise from the mixed vibrations of Li, Zr, and Mg atoms and contribute 48% to the total EPC constant $\lambda$. In contrast, the phonon modes in the intermediate- and high-frequency range (9–40 THz) are largely governed by H-atom vibrations, accounting for 52% of the total EPC constant $\lambda$. The $T_c$ of LiMgZr$_2$H$_{12}$ was evaluated by solving the Allen–Dynes modified McMillan equation [24,29,30]:

$$T_c = \omega_{\log} \frac{f_1 f_2}{1.2} \exp\left(\frac{-1.04(1+\lambda)}{\lambda - \mu^* - 0.62\lambda\mu^*}\right) \qquad (2)$$

where $f_1$ and $f_2$ denote two correction factors, $\omega_{\log}$ denotes the logarithmic average frequency, $\lambda$ is the EPC constant, and $\mu^*$ denotes the effective Coulomb repulsion. The definitions of $\omega_{\log}$ and $\lambda$ are given by:

$$\omega_{\log} = \exp\left(\frac{2}{\lambda}\int\frac{d\omega}{\omega}\alpha^2F(\omega)\ln(\omega)\right) \qquad (3)$$

$$\lambda = 2\int\frac{\alpha^2F(\omega)}{\omega}d\omega \qquad (4)$$

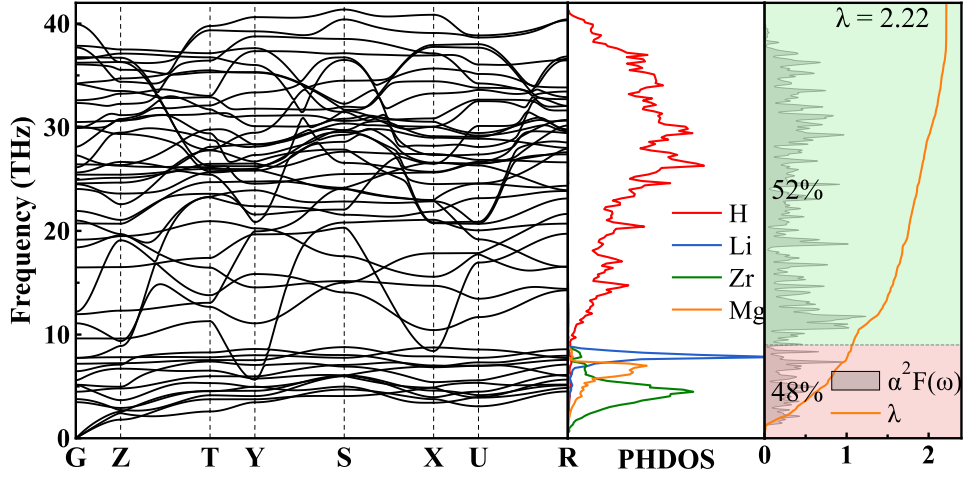

**Fig. 2.** Calculated phonon dispersion curves, phonon density of states, Eliashberg phonon spectral function $\alpha^2F(\omega)$, and EPC intergraded $\lambda$ for LiMgZr$_2$H$_{12}$.

In our calculations, the Coulomb pseudopotential $\mu^*$ was set to 0.10, and the corresponding results are summarized in Table 2. The calculations indicate that LiMgZr$_2$H$_{12}$ exhibits an estimated $T_c$ of 60.8 K at ambient pressure. Table 2 also lists the $T_c$ of MgZrH$_6$ at 36 GPa, as reported by Wang et al. [14], which was calculated using the Allen-Dynes modified McMillan equation. The table demonstrates that Li doping significantly reduces the external pressure required for stabilizing the structure while essentially maintaining its $T_c$.

**Table 2**

EPC constant $\lambda$, logarithmic average phonon frequency $\omega_{\log}$ (in K), density of states at the Fermi level $N_F$ (in States/eV), $T_c$ values (in K) estimated using the Allen-Dynes modified McMillan equation and superconducting figure of merit $S$ of LiMgZr$_2$H$_{12}$ and MgZrH$_6$ under pressures (in GPa).

| Structures | Pressure | $\lambda$ | $\omega_{\log}$ | $N_F$ | $T_c$ | $S$ |
|---|---|---|---|---|---|---|
| LiMgZr$_2$H$_{12}$ | 0 | 2.22 | 396.2 | 2.34 | 60.8 | 1.56 |
| MgZrH$_6$[14] | 36 | 1.13 | | 0.12 | 61.4 | 1.16 |

To better evaluate the overall performance and practical applicability of the superconductor, we calculated the superconducting figure of merit $S$ [31] for LiMgZr$_2$H$_{12}$, which reflects the feasibility of a superconducting material for practical applications. The $S$ parameter is defined as follows:

$$S = T_c / \sqrt{T_{c,MgB_2}^2 + P^2} \tag{5}$$

Here, $T_{c,MgB_2}$ denotes the superconducting critical temperature for MgB$_2$, and $P$ represents the applied pressure. The calculated $S$ value of LiMgZr$_2$H$_{12}$ is 1.56, which is approximately 34% higher than that of the ternary hydrogen-rich compound MgZrH$_6$. In addition, it outperforms a number of other representative, well-known superconducting materials. For example, the $S$ values of experimentally synthesized H$_3$S, YH$_9$, and LaH$_{10}$ have been evaluated as 1.27, 1.19 and 1.43, respectively [1,3,4]; an $S$ value of 1.23 has been reported for the recently synthesized LaBeH$_8$ superconductor [12]. This indicates that LiMgZr$_2$H$_{12}$ has outstanding potential for practical applications.

The electron localization function (ELF) and Bader charges were also calculated, enabling the chemical bonding to be analysed. Fig. 3(a) shows the two-dimensional ELF map of LiMgZr$_2$H$_{12}$. The low ELF values between the metal and hydrogen atoms indicate ionic bonding, consistent with charge transfer from the metal atoms to hydrogen. The ELF value between the nearest neighbor H atoms is approximately 0.45. This low ELF indicates minimal electron localization between H atoms. This confirms the absence of H–H covalent bonds and indicates that hydrogen is stabilized as monatomic species rather than as $H_2$-like units. A subsequent Bader charge analysis shows that each Li, Mg, and Zr atom donates approximately 0.86, 1.63, and 1.66 $e$, respectively, while each H atom gains approximately 0.41–0.55 $e$. These results provide additional evidence for the ionic interaction between the metal and hydrogen atoms in $LiMgZr_2H_{12}$.

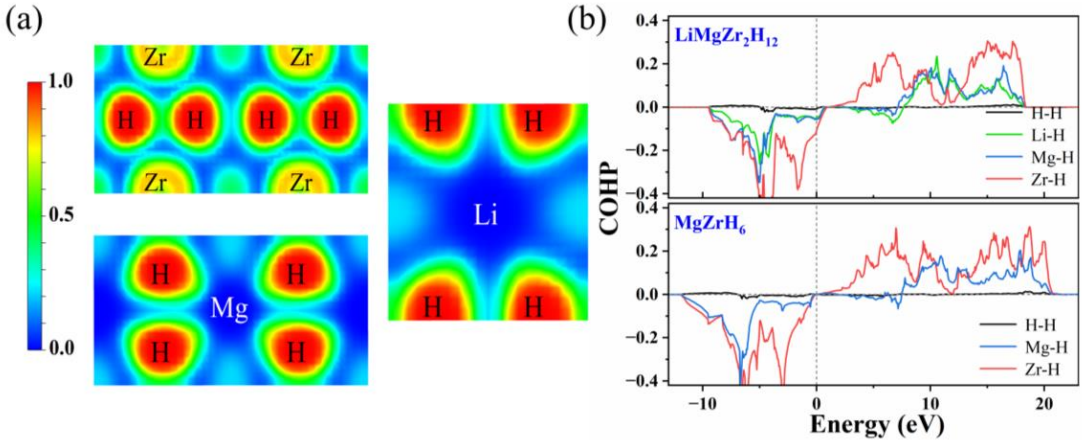

**Fig. 3.** (a) Electronic localization function (ELF) of $LiMgZr_2H_{12}$. (b) The average Calculated crystal orbital Hamiltonian populations (COHP) of selected atomic pairs in $LiMgZr_2H_{12}$ (0 GPa) and $MgZrH_6$ (36 GPa).

Subsequently, to evaluate the interactions between atoms, we calculated the COHP [32] for selected atom pairs in $LiMgZr_2H_{12}$. As shown in Fig. 3(b), pronounced negative peaks appear below the Fermi level for the Zr–H, Mg–H, and Li–H bonds, indicating that bonding is primarily contributed by Zr–H, Mg–H, and Li–H interactions, which play a crucial role in stabilizing the structure. In contrast, the H–H COHP curve remains close to zero over the entire energy range and exhibits only minute oscillations, with almost no discernible peaks. This behavior suggests almost no interactions between neighboring H atoms, which is consistent with the results of the ELF analysis. Moreover, the deepest peak below the Fermi level is associated with the Zr–H bonds, indicating that Zr–H bonding is the strongest in this system. A comparison with the COHP curves of $MgZrH_6$ reveals a key difference: in $LiMgZr_2H_{12}$, the Fermi level lies within a bonding region (negative COHP) dominated by Zr–H bonding states, whereas, in $MgZrH_6$, the Fermi level is located near the boundary between bonding and antibonding states, where the COHP is close to zero. These results indicate that Li doping shifts the Fermi level into a pronounced bonding region, significantly enhancing the electronic density of states at the Fermi level.

The electronic structure of a material is closely related to its superconductivity. Thus, we further investigated the band structures and density of states (DOS) of $LiMgZr_2H_{12}$, as shown in Fig. 4(a). Distinct van Hove singularities appear near the Fermi level at the Y, S, and X points, which can effectively enhance superconductivity. The overall density of states profile of $LiMgZr_2H_{12}$ is similar to that of $Pm\bar{3}$-$MgZrH_6$ at 36 GPa [14]. However, due to the relative downward shift of the Fermi level in $LiMgZr_2H_{12}$, the contribution of H and Zr atoms to the density of states near the Fermi level significantly increases, thereby enhancing the total density of states (TDOS) near the Fermi level. In addition, the density of states near the Fermi level in the $LiMgZr_2H_{12}$ structure is primarily dominated by hydrogen atoms. The contribution of hydrogen near the Fermi level to the TDOS is significantly higher than those of the three metal elements Li, Zr, and Mg. This high hydrogen-derived DOS may enable more electrons to participate in Cooper-pair formation. According to BCS theory [33], a larger TDOS at $E_F$ and H-dominated electronic states at the Fermi level generally favor stronger EPC and a higher $T_c$. This explains why $LiMgZr_2H_{12}$ exhibits superior superconducting properties compared to $MgZrH_6$. Notably, two bands with almost parallel dispersion are found to intersect the Fermi level along the T–Y direction. The presence of such nearly parallel dispersive features in the band structure indicates potential Fermi surface nesting, whose strength is largely governed by the geometry of the Fermi sheets. The Fermi surface topology of $LiMgZr_2H_{12}$ in the Brillouin zone was calculated. As shown in Fig. 4(b), four conduction bands cross the Fermi level in $LiMgZr_2H_{12}$. Here, $n$ denotes the band index. For $n = 1$ and 2, the corresponding Fermi surfaces form closed, smooth, nearly spherical pockets, indicating typical metallic behavior. The $n = 3$ band generates a rhombic closed "inner-shell" Fermi surface, whose four sides contain extended, nearly flat facets that are approximately parallel to the Fermi sheets formed by the $n = 4$ band, resulting in interband nesting channels. Importantly, such nesting may induce phonon softening or Kohn anomalies and enhance EPC, thereby playing a crucial role in strengthening superconductivity [34].

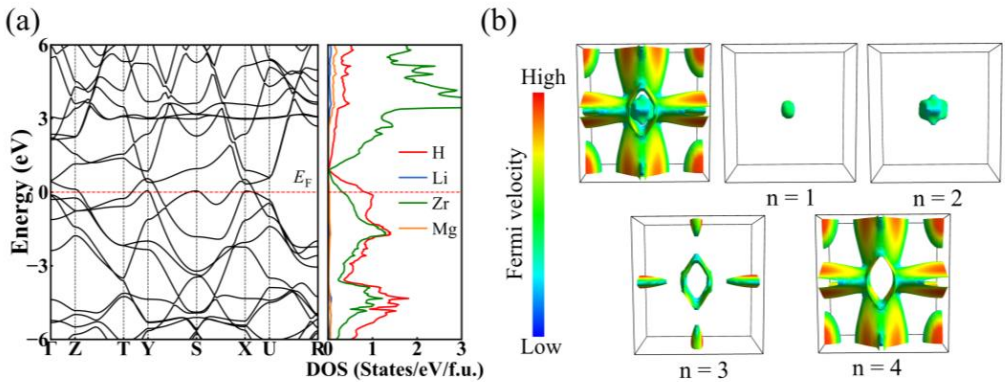

**Fig. 4.** (a) Calculated band structures and PDOS for LiMgZr$_2$H$_{12}$ under ambient pressure. The dashed line at zero indicates the Fermi energy. (b) Fermi surface topology of LiMgZr$_2$H$_{12}$.

Finally, to assess the experimental feasibility of LiMgZr$_2$H$_{12}$, its potential synthesis route was explored. Inspired by previous studies, particularly the work of Yang et al., we explored a possible synthesis route for LiMgZr$_2$H$_{12}$. Yang et al. [35] synthesized a Mg–Zr–Li–H quaternary hydride with a $Fm\bar{3}m$ symmetry at 8 GPa and 873 K via the reaction 6MgH$_2$ + ZrH$_2$ + $n$LiH and found that the formation enthalpy of the quaternary phase is lower than that of Mg–Zr–H ternary hydrides. This result indicates that Mg–Zr–Li–H quaternary hydrides possess better thermodynamic stability. Motivated by this result, we propose a synthesis route for the LiMgZr$_2$H$_{12}$ quaternary hydride: LiMgZr$_2$H$_{12}$ → MgH$_2$ + LiH + 2ZrH$_2$ + 5/2H$_2$. The calculated energy on the left side of this reaction is approximately 170 meV/atom higher than that on the right side, indicating that LiMgZr$_2$H$_{12}$ is a metastable phase. The reaction kinetics under ambient conditions may be slow for such solid metastable phases; thus, catalysts can be used to accelerate the reaction process. In addition, because metastable phases are prone to decomposition or phase transformation during synthesis, measures such as rapid quenching, sealing in quartz ampoules, or handling in an inert gas atmosphere are typically required.

In summary, inspired by recent studies on the MgZrH$_{2n}$ series, we constructed a LiMgZr$_2$H$_{12}$ structure with *Pmmm* symmetry and investigated its stability, electronic properties, and superconductivity using first-principles calculations. The ELF results indicate that there is almost no interaction between H–H pairs in LiMgZr$_2$H$_{12}$ and that the interactions between the metal and hydrogen atoms are predominantly ionic. Furthermore, COHP analysis shows that the Zr–H interaction is the strongest and plays a crucial role in stabilizing the structure. EPC analysis demonstrates that LiMgZr$_2$H$_{12}$ remains dynamically stable at ambient pressure and exhibits a high $T_c$ of 60.8 K. Compared with MgZrH$_6$, Li doping significantly reduces the external pressure required to stabilize the structure while essentially maintaining its $T_c$. This improvement can be attributed to the high electronic DOS at the Fermi level and the strong EPC in LiMgZr$_2$H$_{12}$. H-dominated electronic states at the Fermi level is another key factor that enhances its superconducting performance. Moreover, the calculated superconducting figure of merit $S$ of LiMgZr$_2$H$_{12}$ is 1.56, which is approximately 34% greater than that of MgZrH$_6$ in the ternary system, indicating substantial potential for practical applications. Our study provides theoretical guidance for future experimental work and offers valuable insights into the exploration of quaternary superconducting hydrides, which remain largely unexplored to date.

## Acknowledgements

This research was funded by the National Natural Science Foundation of China (Grant Nos. 11965005 and 11964026), the Natural Science Basic Research Plan in Shaanxi Province of China (Grant Nos. 2025JC-YBMS-027). All the authors thank the computing facilities at High Performance Computing Center of Xidian University.